%% file: main.tex
\documentclass[runningheads]{llncs}
\usepackage{graphicx}
\usepackage{hyperref}

\usepackage{times}
\usepackage{latexsym}
\usepackage{amsmath}
\usepackage{amssymb}
\usepackage{multirow}
\usepackage{url}
\usepackage{booktabs}
\usepackage{pgfplots}
\usepackage{graphicx}
\usepackage{subfig}
\usepackage{wrapfig}
\usepackage{caption}
\usepackage{color}
\usepackage{floatrow}
\usepackage{microtype}
\usepackage{tikz}
\usepackage{listing}
\usepackage{siunitx}
\usepackage{placeins}

\usepackage{lscape}
\usepackage{cleveref}
\usepackage{array}
\usepackage{tabulary}
\newcolumntype{C}[1]{>{\centering\arraybackslash}p{#1}}
\newcolumntype{L}[1]{>{\raggedright\arraybackslash}p{#1}}
\newcolumntype{R}[1]{>{\raggedleft\arraybackslash}p{#1}}

\input{commands.tex}

\begin{document}
\title{Pruning Algorithms for Low-Dimensional Non-Metric k-NN Search: A Case Study\thanks{
Authors gratefully acknowledge the support by the NSF grant \#1618159 : ``Matching and Ranking via Proximity Graphs: Applications to Question Answering and Beyond''}}
%
%
\author{Leonid Boytsov\inst{1}
\and
Eric Nyberg\inst{2}
}
\authorrunning{L. Boytsov and E. Nyberg}
%
\institute{Carnegie Mellon University, Pittsburgh, PA, {srchvrs@cs.cmu.edu}
\and
Carnegie Mellon University, Pittsburgh, PA, {enh@cs.cmu.edu}
}
\maketitle              
\begin{abstract}
We focus on low-dimensional non-metric search,
where tree-based approaches permit efficient and accurate retrieval
while having short indexing time.
These methods rely on space partitioning and require a pruning
rule to avoid visiting unpromising parts.
We consider two known data-driven
approaches to extend these rules
to non-metric spaces: 
TriGen and a piece-wise linear approximation of the pruning rule.
We propose and evaluate two adaptations of TriGen to non-symmetric similarities (TriGen
does not support non-symmetric distances).
We also evaluate a hybrid of TriGen and the piece-wise linear approximation pruning.
We find that this hybrid approach is often more effective
than either of the pruning rules.
We make our software publicly available.
This is an archival version, the publisher's version is available
at \href{https://link.springer.com/chapter/10.1007/978-3-030-32047-8_7}{Springer.com}.
\end{abstract}
\input{intro.tex}

\input{methods_and_data.tex}

\input{exper.tex}

\input{conclusion.tex}

\FloatBarrier


\input{main.bbl}
\end{document}

%% file: commands.tex
\newcommand{\domain}{\ensuremath{D}}
\newcommand{\knn}{\mbox{$k$-NN} }
\newcommand{\topk}{top-\textit{k}}
\newcommand{\ttt}{\texttt}

\newcommand{\knnns}{\mbox{$k$-{NN}}}

\newcommand{\second}[1]{\SI{#1}{\second}}

\newcommand{\normeuc}[1]{\lvert\lvert {#1} \rvert\rvert_2}

\newif\ifProduction

\ifProduction
\newcommand{\leoinline}[1]{}

\else
\newcommand{\leoinline}[1]{\textbf{\color{red}\small #1}}

\fi

%% file: intro.tex
\section{Introduction and Problem Definition}

We consider a $k$ nearest neighbor (\knnns) search,
which is a popular technology used in many domains including, machine learning (ML), data mining, information retrieval, and natural language processing.
Informally, \knn search is a task of retrieving $k$ data set entries closest
to a query point with respect to some distance or similarity function.
This problem originated from the real-world spatial search. 
In particular, Knuth famously formulated \knn search as as the (nearest) post-office problem \cite{knuth1973art}.
With subsequent developments of the vector-space abstraction,
the problem was generalized to searching in a multi-dimensional vector 
and/or generic metric space,
where the latter may lack the structure of the vector space \cite{Samet2005,chavez2001searching}.
Motivated by emergence of useful non-metric distances---such as Bregman divergences \cite{Bregman1967}---the problem was recently generalized to more challenging domains \cite{Cayton2008,DBLP:journals/pvldb/ZhangOPT09,Skopal2011,boytsov2013learning}.

Formally, we assume to have a possibly infinite domain containing
objects $x$, $y$, $z$, \ldots, which are commonly called data points or simply points.
The domain---sometimes called a \emph{space}---is equipped with 
with a \emph{distance function} $d(x,y)$,
which is used to measure dissimilarity of objects $x$ and $y$.
The value of $d(x,y)$ is interpreted as a degree of dissimilarity.
The larger is $d(x,y)$, the more dissimilar points $x$ and $y$ are.
Some distances are non-negative and become zero
only when $x$ and $y$ have the highest possible degree of similarity.
The \emph{metric} distances are additionally symmetric and satisfy the triangle inequality.
However, in general, we do not impose any restrictions on the value of the distance function (except that smaller values represent more similar objects).

We further assume that there is a data set \domain\ containing a \emph{finite}
number of domain points and a set of queries that belong to the domain but not to \domain.
We then consider a standard \topk\  retrieval problem. 
Given a query $q$,
a retrieval task consists in finding $k$ data set points $\{x_i\}$ with smallest values of distances to the query among all data set points (ties are broken arbitrarily). 
Data points $\{x_i\}$ are called \emph{nearest neighbors}.
A search should return $\{x_i\}$ in the order of increasing distance to the query.
If the distance is not symmetric, two types of queries can be considered:
\emph{left} and \emph{right} queries.
In a \emph{left} query, a data point compared to the query is always
the first (i.e., the left) argument of $d(x,y)$.
For simplicity of exposition we consider only the case of left queries.

We employ a space-partitioning method VP-tree \cite{omohundro1989five,Yianilos:1993,uhlmann1991satisfying},
but many other space-partitioning approaches can be used.
Importantly, applying space-partitioning methods to non-metric data of even
moderate dimensionality entails two problems.
First, exact space-partitioning methods can degenerate to a brute-force search 
for just a dozen of dimensions \cite{weber1998quantitative,beyer1999nearest}.
Second, many generic space-partitioning methods
incorporate pruning rules that crucially rely on the triangle inequality,
which does not generally hold in non-metric spaces. 
Most existing non-metric space-partitioning methods employ
specialized extensions specific to a concrete class of distances,
e.g., to  
Bregman divergences \cite{DBLP:journals/pvldb/ZhangOPT09,Cayton2008} 
or Ptolemaic distances \cite{hetland2013ptolemaic}.
However, in a more general case we clearly need to resort to empirically derived
analogs of the triangle inequality,
which are inferred from data with a certain degree of approximation.

\input{table_dist.tex}

\FloatBarrier

For these reasons, we focus only on \emph{approximate} search methods.
We also restrict our attention to low- and moderate-dimensional methods,
because even approximate pruning methods are not effective in truly high dimensions.
There has been a tremendous effort put into design of metric space-partitioning algorithms \cite{chavez2001searching,Samet2005}, 
but many fewer methods are designed for non-metric domains.
We aim to fill this gap by making the following contribution, 
which we detail in the rest of the paper:
\begin{itemize}
    \item We carry out the first experimental comparison of two existing generic pruning algorithms,
    which include the piecewise linear approximation of the pruning rule~\cite{boytsov2013learning} and TriGen \cite{skopal2007}. 
    \item Unlike most prior work, many of our distances are non-symmetric.
    To deal with non-symmetry, we propose two adaptation of TriGen
    to non-symmetric distances and demonstrate that the choice of the symmetrization algorithm can be quite important.
    \item In our comprehensive evaluation, which includes 40 combinations of data sets and distances,
    we demonstrate the feasibility of accurate non-metric \knn search for data of moderate
    dimensionality.
    \item We demonstrate that often best results can be achieved by combining these pruning methods.
    \item We find that on data of moderate dimensionality,  the pruning algorithm needs to be quite efficient.
\end{itemize}

%% file: table_dist.tex
\begin{table}[tb]
\centering
\caption{Distance Functions}
\label{TableDist}
\begin{tabular}{@{}L{5.5cm}l@{\hspace{0.5em}}@{}}
\toprule
Denotation/Name & d(x,y)  \\ \midrule

Euclidean Distance ($L_2$) & $\normeuc{x-y} = \left[\sum\limits_i (x_i-y_i)^2\right]^{1/2}$     \\

$L_p$ ($p>0$) & $\left[\sum\limits_{i=1}^m  (x_i-y_i)^p\right]^{1/p}$ \\

Squared Euclidean ($L^2_2$) & $\normeuc{x-y}^2 = \sum\limits_i (x_i-y^i)^2$  \\

Cosine Distance & $1-\dfrac{\sum_i x_i y_i}{\normeuc{x}\normeuc{y}}$  \\\midrule

Kullback-Leibler diverg. (KL-div.) \cite{kullback1951} & $\sum\limits_{i=1}^m  x_i\log{\dfrac{x_i}{y_i}}$ 
\\

Itakura-Saito distance \cite{itakura1968analysis} & $\sum\limits_{i=1}^m \left[ \frac{ x_i}{y_i} - \log \frac{x_i}{y_i}  -1 \right]$ 
\\

R\'{e}nyi diverg. \cite{rrnyi1961measures} &  $\frac{1}{\alpha-1}\log\left[\sum\limits_{i=1}^m x_i^\alpha y_i^{1-\alpha}\right]$,\;
$\alpha > 0$ and $\alpha \ne 0$
\\
\bottomrule
\end{tabular}
\end{table}

%% file: methods_and_data.tex
\section{Methods and Materials}
\subsection{Data sets and Distances}\label{SectionData}

In our  experiments, we use the following non-metric distances:
$L_2^2$ (squared Euclidean)
$L_p$ distance,
cosine distance,
KL-divergence, 
the Itakura-Saito distance,
and the family of R\'{e}nyi divergence distances.
The first three distances are symmetric.
The remaining distances are statistical distances defined
over probability distributions.
For expository purposes, we also use the Euclidean metric distance $L_2$.
Distances are listed in Table~\ref{TableDist}.

Statistical distances in general and, KL divergence in particular, play an important role in ML \cite{Cayton2008,markatou2017statistical}. 
They are typically non-symmetric.
Both the KL-divergence and the Itakura-Saito distances
were used in prior work \cite{Cayton2008}.
The R\'{e}nyi divergence is a single-parameter family of distances,
which are not symmetric when the parameter $\alpha \ne 0.5$.
By changing the parameter  we can vary the degree of symmetry.
In particular, large values of $\alpha$ and close-to-zero values
result in highly non-symmetric distances.
This flexibility allows us to ``stress-test'' retrieval methods
on challenging non-symmetric distances.

\input{table_data.tex}

The data sets are listed in Table~\ref{TableData}.
Wiki-$d$ and RCV-$d$
data sets consist of dense vectors of topic histograms with $d$ topics.
RCV-$d$ set are created by Cayton \cite{Cayton2008}  by applying the latent Dirichlet allocation (LDA) method \cite{blei2003latent} 
to the RCV1 collection \cite{DBLP:journals/jmlr/LewisYRL04}. 
These data sets have only 500K entries.
Thus, we created 
larger sets from Wikipedia following a similar methodology.
RandHist-$d$ is a synthetic set of topics sampled uniformly 
from a $d$-dimensional simplex.

\subsection{Pruning Algorithms for Space-Partitioning Methods}\label{SectionMethods}
We employ a simple approach called a \emph{vantage-point} tree (VP-tree) \cite{omohundro1989five,Yianilos:1993,uhlmann1991satisfying}.
There are two reasons for this choice: 
for low- and moderate-dimensional data, it is often a hard-to-beat method. 
For example, in a  preliminary experiment with $L_2$ on Wiki-8 data set
for exact 10-NN search using NMSLIB \cite{SISAP2013}, 
 SA-tree \cite{Navarro2002}, 
GH-tree \cite{uhlmann1991satisfying}, 
MVP-tree (binary version) \cite{DBLP:journals/tods/BozkayaO99}, 
and VP-tree are respectively
70$\times$, 210$\times$, 1200$\times$,  1600$\times$
faster than the brute-force search.
This comparison was done using the leaf bucket of size 50 for all methods (except SA-tree, which does not easily support bucketing)
and without using any specific optimizations for any of the methods.
We can see that VP-tree can outperform fancier alternatives including MVP-tree, which 
carries out 3$\times$ fewer distance computations in this experiment.

VP-tree is a hierarchical space-partitioning method, which divides the space using hyperspheres.
The output of an indexing algorithm is a hierarchical partitioning of the data set  represented by a binary tree.
This algorithm is a \emph{recursive} procedure 
 that operates on a subset of data---which we call an \emph{active subset}---and on a partially built tree.
 At each step of recursion, the indexing algorithm checks if the number of active data points is below a certain threshold called the \emph{bucket} size.
If this is the case, the active data points are simply stored as a single bucket.
Otherwise, the algorithm divides the active subset into two nearly equal parts,
each of which is further processed recursively.

Division of the active subset starts with selecting a pivot $\pi$ (e.g., randomly)
and computing the distance from $\pi$ to every other data point in the active subset.
Assume that $R$ is the median distance.
Then, the active subset is divided into two subsets by the hypersphere with radius $R$ and center $\pi$.
Two subtrees are created.
Points inside the pivot-centered hypersphere are placed into the left subtree.
Points outside the pivot-centered hypersphere are placed into the right subtree.
Points on the separating hypersphere may be placed arbitrarily.
Because $R$ is the median distance, each of the subtrees contains approximately half of active  points.

In VP-tree \knn search can be seen as a range search with a shrinking radius.
The search algorithm is a \emph{best-first} traversal procedure that starts from the root of the tree and proceeds recursively.
It updates the search radius $r$ as it encounters new close data points.
Let us consider one step of recursion.
If the search algorithm reaches a leaf of the tree, i.e., a bucket,
all bucket elements are compared against the query. In other words, elements in the buckets
are searched via brute-force search.

If the algorithm reaches an internal node $X$, 
there are exactly two subtrees representing two spaces partitions. 
The query belongs to exactly one partition.
This is the ``best'' partition and the search algorithm always explores this partition recursively
before deciding whether to explore the other partition.
While exploring the best partition, we may encounter new close data points (pivots or bucket points) and further shrink the search radius.
On completing the sub-recursion and returning to node $X$, we make a decision about pruning or exploring the other partition.

\input{query_bal_types.tex}

\paragraph{Piecewise-linear Approximation of the Decision Rule}
An essential part of this process is a decision function,
which identifies situations when pruning is possible without sacrificing accuracy.
Let us review the decision process.
Recall that each internal node keeps pivot $\pi$ and radius $R$,
which define the division of the space into two subspaces.
Although there are many ways to place a query ball,
all locations can be divided into three categories,
which are illustrated by Figure~\ref{FigQueryBallTypes}.
The red query ball ``sits'' inside the inner partition.
Note that it does not intersect
with the outer partition. For this reason, the outer partition cannot have sufficiently
close data points, i.e., points with radius $r$ from the query.
Hence, this partition can be safely pruned.
The blue query ball is located in the outer partition. Likewise,
it does not intersect the other, inner, partition. Thus, this inner partition can be safely pruned.
Finally, the gray query ball intersects both partitions.
In this situation, sufficiently close points may be located in both partitions and no safe pruning is possible.

\input{table_dec_func.tex}

The pruning algorithm can be seen as the \emph{binary classification problem},
which tells us whether we should visit both partitions or only the partition
that contains the query. 
As we show previously ~\cite{boytsov2013learning},
the problem can be solved by collecting training data and building
a non-parametric model,
but a simple two-parameter approach---described below---delivers better results.
Let us first consider the case of a metric distance.
From the triangle inequality it follows  
that the VP-tree search algorithm
visits:
\begin{itemize}
\item \emph{only} the left subtree    if $d(\pi, q) < R - r$;
\item \emph{only} the right subtree   if $d(\pi, q) > R + r$;
\item both subtrees if                $R - r \le d(\pi,q) \le R + r$.
\end{itemize}
Let us rewrite these rules using notation $D_{\pi,R}(x) = |R-x|$.
It is easy to see that the search algorithm has to visit both partitions if and only
if $r \ge D_{\pi,R}(d(\pi, q))$.
If $r < D_{\pi,R}(d(\pi, q))$, we need to visit only one partition that contains the query point whereas the other partition can be safely pruned.

In other words, 
the pruning decision is made
by comparing the query radius $r$ with the value of the function $D_{\pi,R}(x)$, 
whose only argument is the distance from the query to the pivot $d(\pi, q)$.\footnote{
Recall that \knn search is executed as a best-first range search with a shrinking radius}
This basic rule can also be learned from data for non-metric distances.
Our initial approach to learn $D_{\pi,R}(x)$  employed a stratified sampling procedure (see \S~2
of the supplemental materials of our publication \cite{boytsov2013learning}).
However, it was expensive and not very accurate. For this reason, we also implemented a simple \emph{parametric} approximation
whose parameters are selected to optimize efficiency at a given value of recall.

To choose the right parametric representation, 
we examine the (\emph{approximate}) functions $D_{\pi,R}(x)$ learned by the sampling algorithm. 
Plots of functions $D_{\pi,R}(x)$ learned from data are shown in Fig.~\ref{FigDecisionFunc}.
Small dots in these plots represent function values obtained by sampling.
Blue curves are fit to these dots.
In these plots, we use only topic histogram data RCV-$d$, where $d \in \{8, 32\}$
and random 8-dimensional histograms (RandHist-8).

For the Euclidean data (Panels \ref{PanelDecGraph_RCV8_L2}-\ref{PanelDecGraph_RandHist8_L2} in Figure \ref{FigDecisionFunc}),
$D_{\pi,R}(x)$ resembles a piecewise linear function close to the exact metric 
pruning function $|R-x|$.
For the KL-divergence data (Panels~\ref{PanelDecGraph_RCV8_KLdiv}-\ref{PanelDecGraph_RandHist8_KLdiv} in Figure \ref{FigDecisionFunc}),
$D_{\pi,R}(x)$ looks like either a U-shape or a hockey-stick curve.
These observations originally motivated the use of a piecewise \emph{polynomial} decision function, which is formally defined as:
\begin{equation}\label{EqDecFuncPolynom}
D_{\pi,R}(x) = \left\{
\begin{array}{ll}
\alpha_{left}  |x - R|^{\beta_{left}},  & \mbox{ if }x \le R\\
\alpha_{right} |x - R|^{\beta_{right}}, & \mbox{ if }x \ge R\\
\end{array}
\right.,
\end{equation}
where $\beta_{i}$ are positive integers.
However, preliminary experiments convinced us to switch to a simple piece-wise linear variant.
First, we learned that using different $\beta_i$ did not make our pruning function
sufficiently more accurate. However, it made the optimization problem harder
due to additional parameters (so we set $\beta = \beta_1 = \beta_2$).
\label{LinearVsPolynomDecisionDisc}
Second, we found that in many cases a polynomial approximation was not better than a piecewise linear one, especially when dimensionality was high.

This is not very surprising: Due to the concentration of measure,
for most data points the distance to the pivot $\pi$ is close to the median distance $R$
(which corresponds to the boundary between two VP-tree partitions).
If we explore the shape of $D_{\pi,R}(x)$ in Panels~\ref{PanelDecGraph_RCV8_L2}
and \ref{PanelDecGraph_RCV32_KLdiv}  around the median,
we can see that a  piecewise linear shape approximation is quite reasonable.
To sum up, we ended up using the piecewise linear parametric decision rule defined as:
\begin{equation}\label{EqDecFuncLinear}
D_{\pi,R}(x) = \left\{
\begin{array}{ll}
\alpha_{left} |x - R|,  & \mbox{ if }x \le R\\
\alpha_{right} |x - R|, & \mbox{ if }x \ge R\\
\end{array}
\right.
\end{equation}
This is similar to stretching of the triangle inequality proposed
by Ch\'{a}vez and Navarro~\cite{Chavez_and_Navarro:2003}.
There are two crucial differences, however. First, we utilize different values of $\alpha_i$, i.e., $\alpha_{left} \ne \alpha_{right}$,
while  Ch\'{a}vez and Navarro used $\alpha_{left} = \alpha_{right}$.
Second, we devise a simple training procedure to obtain values of $\alpha_i$ that maximize efficiency at a given recall value.
For details, we address the reader to relevant publications \cite{boytsov2013learning,boytsov2018efficient}.

\paragraph{TriGen}
TriGen consists in ``stretching'' the distance function using a monotonic
concave transformation \cite{skopal2007} that reduces non-metricity of the distance.
TriGen is designed only for \emph{bounded}, \emph{semimetric} distances,
which are crucially \emph{symmetric}, non-negative, and become zero only for identical data points. 
We are not aware of any prior extensions to non-symmetric distances except 
a straightforward filter-and-refine approach. 

Let $x$, $y$, $z$ be an arbitrary ordered triple of points such
that $d(x,y)$ is the largest among three pairwise distances,
i.e., $d(x,y) \ge \max(d(x,z),d(z,y))$.
If $d(x,y)$ is a metric distance, the following conditions should all be true:
\begin{equation}\label{EqTripleIneq}
\begin{array}{c}
d(x,y) \le d(x,z) + d(z,y)
\\
d(y,z) \le d(y,x) + d(x,z)
\\
d(x,z) \le d(x,y) + d(y,z)
\end{array}
\end{equation}
Because $d(x,y) \ge \max(d(x,z),d(z,y))$, the second and the third
inequalities in (\ref{EqTripleIneq}) are trivially satisfied for (not necessarily metric) \emph{symmetric}
and \emph{non-negative} distances.
However, the first condition can be violated if the distance is non-metric.
The closer is the distance to the metric distance, the less frequently we encounter such violations.
Thus, it is quite reasonable
to assess the degree of deviation from metricity
by estimating a probability that the triangle inequality is violated (for a randomly selected triple),
which is exactly what is suggested by Skopal \cite{skopal2007}.

Skopal proposes a clever way to decrease non-metricity by
constructing a new distance $f(d(x,y))$, where $f()$ is a monotonically increasing concave function.
The concave function ``stretches'' the distance and makes it more similar to a true metric compared to the original distance $d(x,y)$.
At the same time, due to the monotonicity of such a transformation,
the \knn search using the modified distance
produces the same result as the \knn search using the original distance.
Thus, the TriGen strategy to dealing with non-metric data consists in (1)
employing a monotonic transformation that makes a distance approximately metric while preserving
the original set of nearest neighbors, and (2)
indexing data using an exact metric-space access method.

A TriGen mapping $f(x)$---defined  for $0 \le x \le 1$---is selected from the union
of two parametric families of concave functions,
which are termed as bases:
\begin{itemize}
\item A fractional power base $FP(x,w)=x^\frac{1}{1+w}$;
\item A Rational B\'{e}zier Quadratic (RBQ) base
$RBQ_{(a,b)}(x,w)$, $0 \le a < b \le 1$.
The exact functional form of RBQ is not relevant to this discussion (see \cite{skopal2007} for details).
\end{itemize}
Note that parameters $w$, $a$, and $b$ are treated as constants,
which define a specific functional form.
By varying these parameters we can design a necessary stretching function.
The larger is the value of $w$ the more concave is the transformation and the more ``metric''
is the transformed distance.
In particular, as $w \rightarrow \infty$, both RBQ and FP converge to
one minus the Dirac delta function.
This limit function of all bases is
equal to zero for $x=0$ and to one for $0 < x \le 1$.
As noted by Skopal \cite{skopal2007}, applying such a degenerate transformation
produces a \emph{trivial} metric space where $d(x,x)=0$ and $d(x,y)=C$ 
for some constant $C>0$ and all $x \ne y$.

A learning objective of TriGen, however,
is to select a \emph{single} concave function that
satisfies the accuracy requirements
while allowing for efficient retrieval.
The fraction of violations is computed for a set of \ttt{trigenSampleTripletQty}
ordered data point triples sampled from a set of \ttt{trigenSampleQty} data points,
which are, in turn, selected randomly from the data set (uniformly and without replacement).
The fraction of violations is required to be above the threshold \ttt{trigenAcc}.
Values \ttt{trigenSampleTripletQty}, \ttt{trigenSampleQty}, and \ttt{trigenAcc} are
all parameters in our implementation of TriGen.
To assess efficiency Skopal uses
the value of an intrinsic dimensionality as a proxy metric (see \cite{skopal2007} for details).
The idea is that the modification of the distance with the lowest intrinsic
dimensionality should result in the fast retrieval method.

Because it is not feasible to optimize over the infinite set of transformation functions,
TriGen employs a finite pool of bases,
which includes FB and multiple RBQ bases for all possible combinations of parameters $a$ and $b$
such that $0 \le a < b \le 1$.
For each base, TriGen uses a binary search to find the minimum parameter $w$ such
that the transformed distance deviates from a metric distance within specified limits.
Then the base with minimum intrinsic dimensionality is selected.

TriGen has two major limitations: In addition to be non-negative, the distance should
be symmetric and bounded.
Bounding can be  provided by using $\min(d(x,y)/D_{\textrm{max}}, 1)$ instead of the original distance.\footnote{For efficiency reasons this is simulated via multiplication by inverse maximum distance.}
Note that $D_{\textrm{max}}$  is an empirically estimated maximum distance (obtained by computing $d(x,y)$ for a sample
of data set point pairs).

As noted by Skopal \cite{skopal2007}, searching with a non-symmetric distance can be partially
provided by a filter-and-refine approach
where a fully \emph{min}-symmetrized distance $\min(d(x,y),d(y,x))$
is used during the filtering step.
However, as we learn from our prior work \S\S~2.3.2.3-2.3.2.4
 \cite{boytsov2018efficient},
the filtering step has to carry out a $k_c$-NN search
with $k_c$ (sometimes substantially) larger than $k$.
This is required to compensate for the lack of accuracy due to replacing
the original distance with the symmetrized one.
In that, using $k_c > k$ leads to reduced efficiency.
Thus, instead of the complete filter-and-refine symmetrization, 
we consider two simple alternatives.
In both cases we first apply the TriGen algorithm to the 
min-symmetrized distance.
As a result, we obtain a mapping that
makes this min-symmetrized distance to be closer to  a metric distance.
However, this mapping is used differently in the two modifications of TriGen.

Recall that in a typical space-partitioning method,
we divide the data into reasonably large buckets (50 in our experiments).
The \knn search is simulated as a range search with a shrinking radius.
In the case the first modification of TriGen,
while we traverse the tree, we compute the original and the min-symmetrized distance for two purposes:
\begin{itemize}
\item shrinking the dynamic radius of the query using the \emph{symmetrized} distance;
\item checking if the \emph{original} distance is small enough to update the current set of $k$ nearest neighbors.
\end{itemize}
When we reach a bucket, for every data point in the bucket,
we can compute both the original and the symmetrized distance.
The symmetrized distance is used to update the query radius, while the original distance is used
to update the set of $k$ nearest neighbors.
This is our first modification of TriGen which we refer to as \emph{TriGen 0}.

In the second variant of TriGen, which we refer to as \emph{TriGen 1},
we use \emph{only} the original distance to compute the distance
from the query to bucket data points.
When we compute the distance to the pivots, we  compute
the min-symmetrized distance and apply a metrizing transformation.
However, when we process bucket data points,
we compute only the original distance.
Consequently,
we shrink the dynamic query radius using values of $f(d(x,y))$ instead of
$\min\left(f(d(x,y)), f(d(y,x))\right)$,
In TriGen 1, we expect the query radius to shrink somewhat slower compared to TriGen 0,
which, in turn, can reduce the effectiveness of pruning.
However, we hope that nearly halving the number of distance computations would have a larger effect on overall retrieval time.

%% file: table_data.tex
\begin{table}[tb]
\centering
\caption{Data sets\label{TableData}}
\begin{tabular}{l@{\hspace{1em}}l@{\hspace{1em}}l@{\hspace{1em}}L{6cm}}
\toprule
Name & max. \# of rec. & Dimensionality   & Source   \\ \midrule

RandHist-$d$  & $0.5\times10^6$  & $d = 8$ & Histograms sampled uniformly from a simplex   \\
RCV-$d$ & $0.5\times10^6$ & $d \in \{8, 32, 128\}$  & $d$-topic LDA \cite{blei2003latent} RCV1 \cite{DBLP:journals/jmlr/LewisYRL04} histograms \\
Wiki-$d$ & $2\times10^6$  & $d \in \{8, 32, 128\}$  & $d$-topic LDA \cite{blei2003latent} Wikipedia histograms  \\
\\ \bottomrule
\end{tabular}
\end{table}

%% file: query_bal_types.tex
\begin{wrapfigure}{r}{0.28\textwidth}
\centering
\includegraphics[scale=0.9]{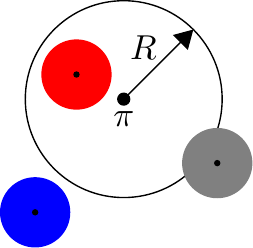}
\caption{Three types of query balls in VP-tree. \label{FigQueryBallTypes}}
\end{wrapfigure}

%% file: table_dec_func.tex
\begin{figure}[tb]
\centering
\subfloat[\scriptsize\label{PanelDecGraph_RCV8_L2}RCV-8 ($L_2$)]{\includegraphics[width=0.320000\textwidth]{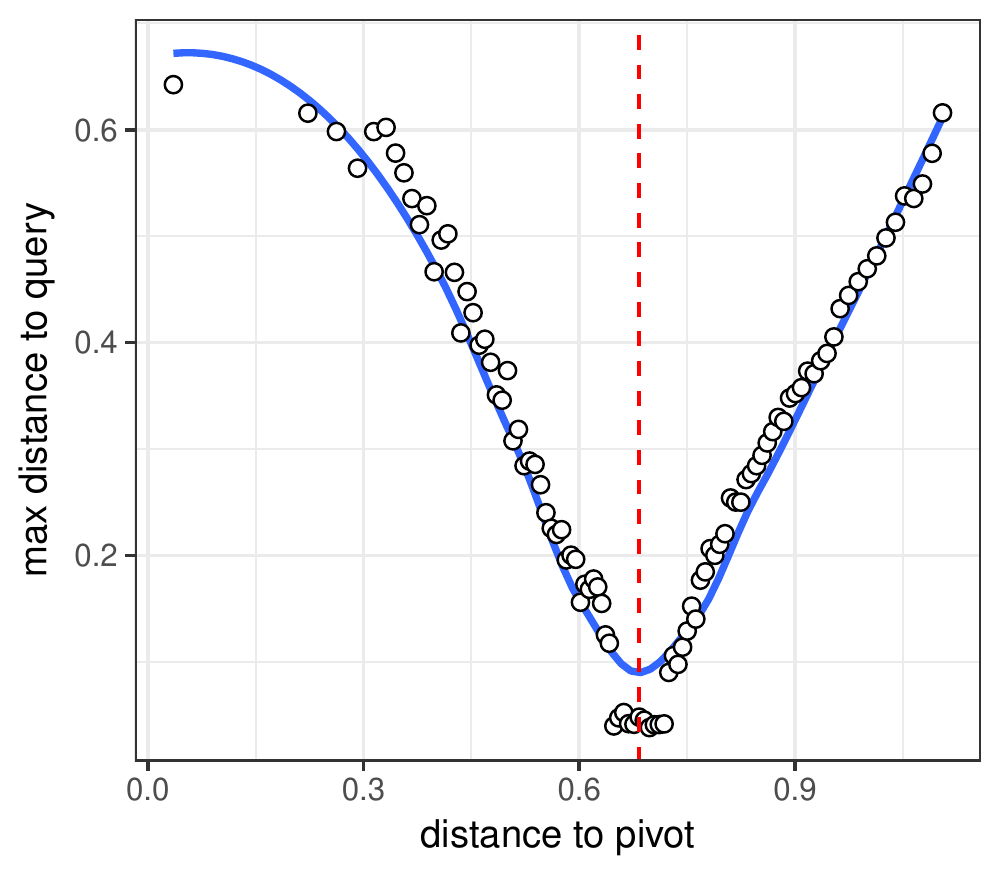}}
\subfloat[\scriptsize\label{PanelDecGraph_RCV32_L2}RCV-32 ($L_2$)]{\includegraphics[width=0.320000\textwidth]{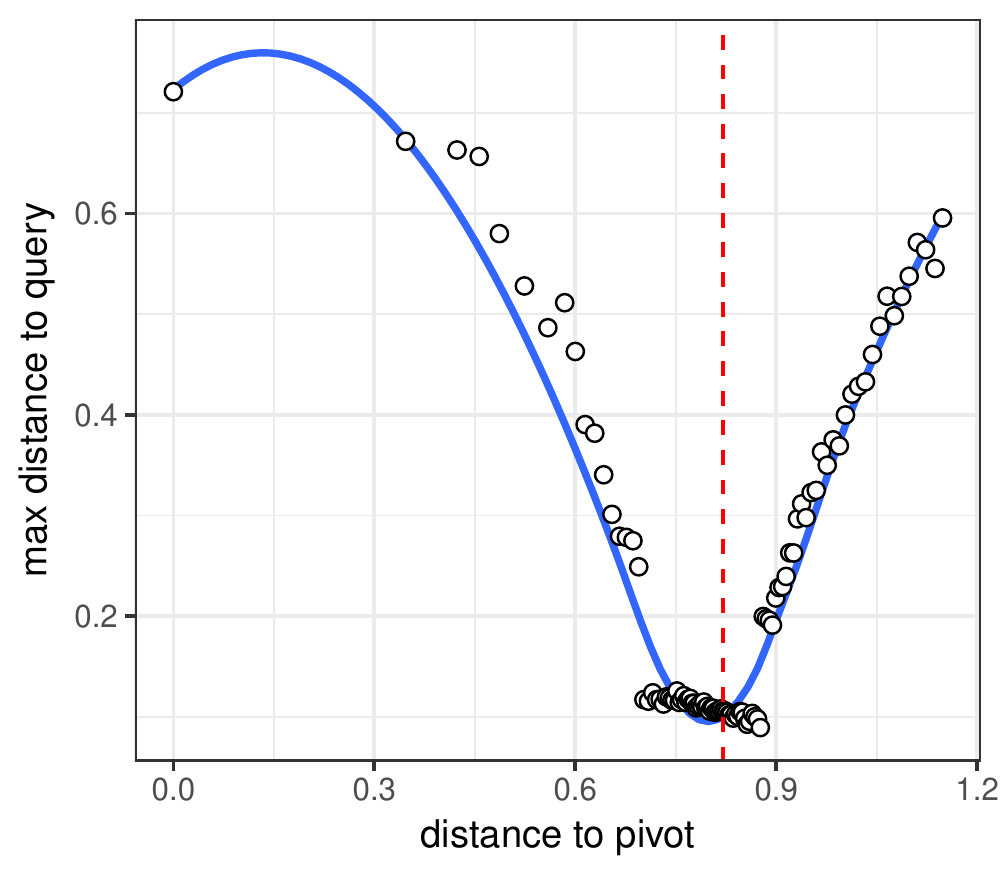}}
\subfloat[\scriptsize\label{PanelDecGraph_RandHist8_L2}RandHist-8 ($L_2$)]{\includegraphics[width=0.320000\textwidth]{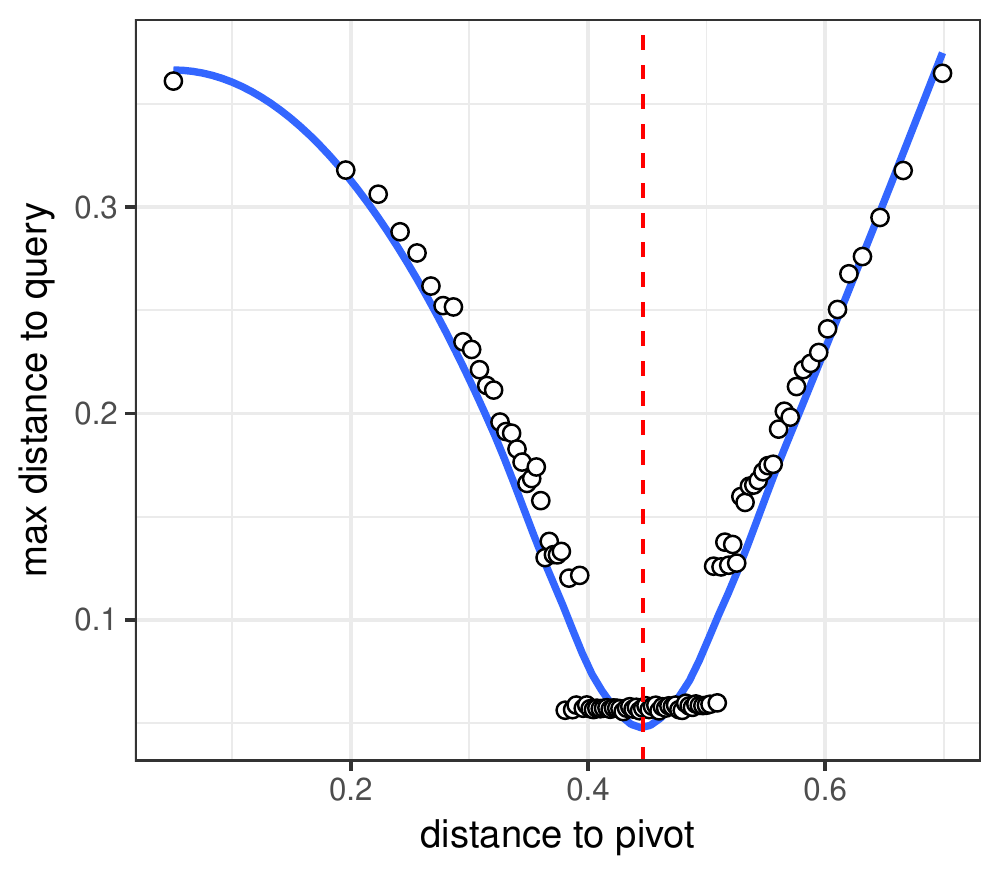}}
\\
\subfloat[\scriptsize\label{PanelDecGraph_RCV8_KLdiv}RCV-8 (KL-div.)]{\includegraphics[width=0.320000\textwidth]{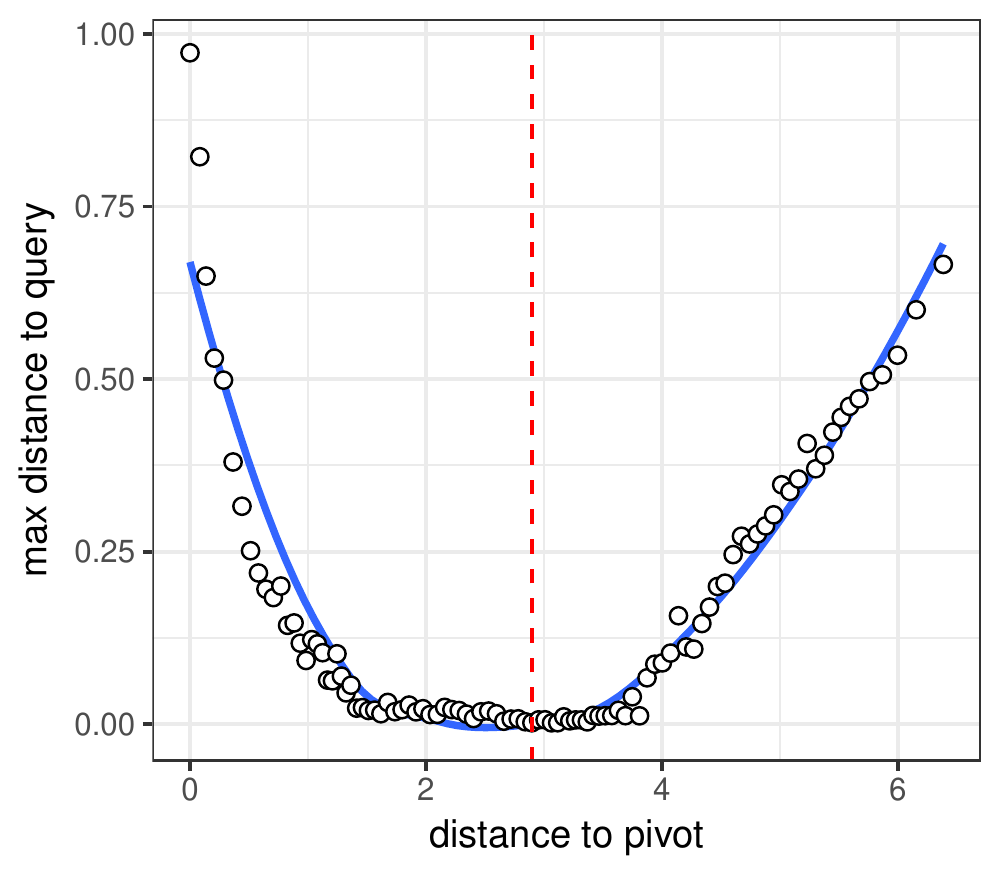}}
\subfloat[\scriptsize\label{PanelDecGraph_RCV32_KLdiv}RCV-32 (KL-div.)]{\includegraphics[width=0.320000\textwidth]{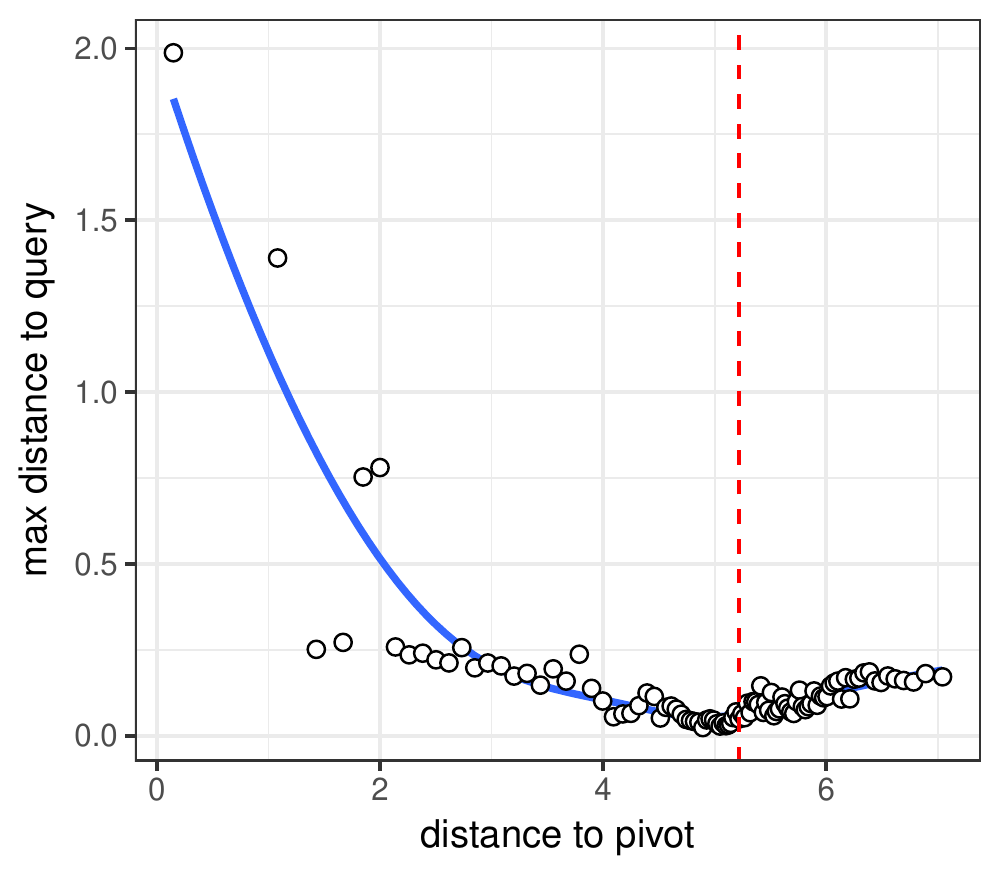}}
\subfloat[\scriptsize\label{PanelDecGraph_RandHist8_KLdiv}RandHist-8 (KL-div.)]{\includegraphics[width=0.320000\textwidth]{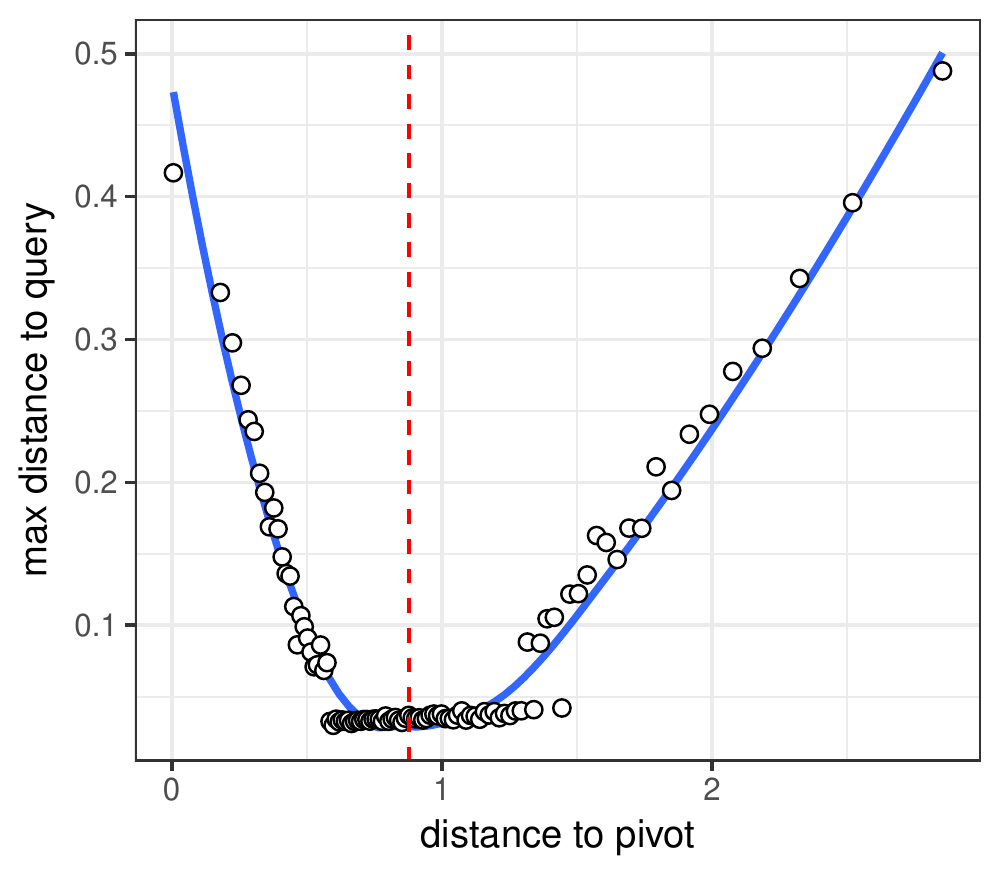}}
\caption{\label{FigDecisionFunc} The empirically obtained (\emph{approximate}) pruning decision function  $D_{\pi,R}(x)$}
\end{figure}

%% file: exper.tex
\input{table_exper_vp.tex}

\section{Experiments}
\subsection{Experimental Setup and Preliminary Experiments}

We compare TriGen and the piecewise-linear pruning approach using the NMSLIB \cite{SISAP2013} implementation
of the VP-tree (method \ttt{vptree\_trigen})\footnote{ \url{https://github.com/nmslib/nmslib/tree/nmslib4a_bigger_reruns}}).
 Experiments are run on a laptop (i7-4700MQ @ 2.40GHz with 16GB of memory).
The accuracy of retrieval is measured via recall (equal to the 
average fraction of neighbors found).

\FloatBarrier
\input{table_trigen.tex}

\FloatBarrier

We use two variants of TriGen ({TriGen 0} and {TriGen 1}),
but for symmetric distances, we use only TriGen 1.
The TriGen algorithm that finds an optimal mapping function was
downloaded from the author's website\footnote{\url{http://siret.ms.mff.cuni.cz/skopal/download.htm}} and incorporated into NMSLIB.
The optimization procedure employs
a combination of parameters $a$ and $b$,
where $a$ are multiples of 0.01, $b$ are multiples of 0.05, and $0 \le a < b \le 1$.
The sampling parameters are set as follows:
\ttt{trigenSampleTripletQty}=10000, \ttt{trigenSampleQty}=5000.

TriGen is compared against two variants of NMSLIB VP-tree,
which rely on the piecewise-linear pruner.
The second variant
uses a clever TriGen idea of applying a concave mapping to make the distance more similar to a metric one. 
However, 
unlike TriGen \cite{skopal2007},
we do not carry an extensive search for an optimal transformation
but rather apply, perhaps, the simplest and fastest
\emph{monotonic concave} transformation possible, which consists in taking a square root. 
On Intel the square root is computed the instruction
\ttt{sqrtss}, which typically takes less than 10 cycles \cite{fog2011instruction}.

In our main experiments,
 we employ 40 combinations of data sets and distances.
All distances are non-metric: We experiment with both symmetric and non-symmetric ones.
Due to space limitations, we do not present all the results
here and refer the reader to our unpublished
technical report for the complete set of results (\S 2.3.3 \cite{boytsov2018efficient}).

Before we proceed, we must answer the following question: ``How difficult are these data sets and distances''? To ensure
we do not deal with mildly non-metric data,
we attempted to index this data using a metric variant
VP-tree without adapting the pruning rule to non-metric distances.
Results for randomly selected 1K queries are presented in Table~\ref{TableMetrVptree}  (for
a subset of distances and data sets),
where we show improvement in efficiency and respective recall.

It can be seen that nearly all the combinations of data and distance functions are substantially non-metric:
Searching using a metric VP-tree is usually fast, but the accuracy is \emph{unacceptably} low.
In particular, this is true for Wiki-8 and Wiki-128 data sets with KL-divergence (which are also used in our main experiments).
One exception, is the $L_p$ distance for $p=0.5$, where recall of about 90\% is achieved for three low-dimensional data sets.
However, as $p$ decreases, the recall decreases sharply, i.e., the distance function becomes ``less'' metric.
To summarize, we clearly deal with challenging non-metric data sets, where both accurate and efficient retrieval
is not possible to achieve by a straightforward application of metric-space search methods.

\subsection{Main Experiments}
Experimental results for 16 out of 40 cases are presented in 
Figures~\ref{FigTrigen_eff} and \ref{FigTrigen_dist}.
The remaining results can be found in the technical report (\S 2.3.3 \cite{boytsov2018efficient}).
In Figure \ref{FigTrigen_eff},
we measure efficiency directly in terms of wall-clock time improvement over the brute-force search.
In Figure \ref{FigTrigen_dist}, we show improvement in the number of distance computations (again
compared to the brute-force search).

First and foremost, 
we can see that VP-tree with a data-adapted
pruning rule can enable accurate non-metric \knn search for data of moderate dimensionality.
When comparing TriGen against the piecewise linear pruner in terms
of pure efficiency, the results are a bit of the mixed bag.
Yet, the piecewise linear pruner is typically better  
(in 23 cases out of 40 on the full set, see \S~2.3.3~\cite{boytsov2018efficient}).

However, the piecewise linear pruner combined with the square-root distance transform
is nearly always better than the basic piecewise linear pruner.
In Panels \ref{PanelTrigen_eff_Wiki8_Cosine},\ref{PanelTrigen_eff_Wiki8_L2SQR}, \ref{PanelTrigen_eff_RCV8_KLdiv},
 \ref{PanelTrigen_eff_RCV8_RenyiDiv025}, \ref{PanelTrigen_eff_RCV8_RenyiDiv075}
the improvement is up to one order of magnitude.
The combination of the piecewise linear pruner with the square root transform outperforms
TriGen in all but two cases, sometimes by an order of magnitude.
In Panels~\ref{PanelTrigen_eff_Wiki8_KLdiv} and \ref{PanelTrigen_eff_Wiki8_ItakuraSaito}, however, TriGen can also be an order of magnitude faster than the piecewise linear pruner.

It is important to note, however,
that there is often little to no difference between 
the hybrid pruning approach and TriGen in terms of the reduction
in the number of distance computations (see Table~\ref{FigTrigen_dist}).
The most likely explanation for this discrepancy is that the transformation functions used in the adopted TriGen implementation are quite expensive to compute.

Finally, we can see that TriGen 1 is never less efficient than TriGen 0.
Furthermore, TriGen 1 is up two times more efficient in four cases (see
Panels~\ref{PanelTrigen_eff_Wiki8_RenyiDiv025},\ref{PanelTrigen_eff_Wiki8_RenyiDiv075},\ref{PanelTrigen_eff_Wiki128_RenyiDiv025},\ref{PanelTrigen_eff_Wiki128_RenyiDiv2}).
This is somewhat unsurprising, because TriGen 0 computes both $d(x, q)$ and
$d(q,x)$ for every data point visited by the search. 
Although this may permit a more effective pruning, 
the cost of extra distance computations outweigh the benefits (at least on our data).


%% file: table_exper_vp.tex
\begin{table}[tb]
\centering
\caption{Efficiency-effectiveness results for metric VP-tree on non-metric data for $10$-NN search (using complete data sets).}
\label{TableMetrVptree}
\begin{tabular}{lcC{1cm}cC{1cm}cC{1cm}cC{1cm}}
\toprule
& \multicolumn{2}{c}{RCV-8} & \multicolumn{2}{c}{Wiki-8} & \multicolumn{2}{c}{RandHist-8} & \multicolumn{2}{c}{Wiki-128} \\\midrule
& Recall & Impr. in eff. & Recall & Impr. in eff. & Recall & Impr. in eff. & Recall & Impr. in eff. \\\midrule
$L_p (p=0.125)$ & 0.41 & 1065 & 0.66 & 15799 & 0.45 & 136 & 0.07 & 14845 \\
$L_p (p=0.25)$ & 0.61 & 517 & 0.78 & 14364 & 0.66 & 115 & 0.09 & 396 \\
$L_p (p=0.5)$ & 0.91 & 926 & 0.94 & 14296 & 0.92 & 174 & 0.50 & 33 \\
$L^2_2$ & 0.69 & 1607 & 0.78 & 5605 & 0.56 & 1261 & 0.55 & 114 \\
Cosine dist. & 0.67 & 1825 & 0.62 & 3503 & 0.58 & 758 & 0.73 & 55 \\
R\'{e}nyi div. ($\alpha=0.25$) & 0.66 & 5096 & 0.70 & 24246 & 0.50 & 3048 & 0.48 & 1277 \\
R\'{e}nyi div. ($\alpha=0.75$) & 0.61 & 9587 & 0.66 & 35940 & 0.50 & 4673 & 0.50 & 468 \\
R\'{e}nyi div. ($\alpha=2$) & 0.40 & 22777 & 0.66 & 46122 & 0.38 & 11762 & 0.71 & 55 \\
KL-div. & 0.52 & 1639 & 0.67 & 5271 & 0.46 & 610 & 0.56 & 41 \\
Itakura-Saito & 0.46 & 706 & 0.69 & 4434 & 0.41 & 1172 & 0.14 & 384 \\
\bottomrule
\end{tabular}
\end{table}

%% file: table_trigen.tex
%
%
%

\begin{figure}[t]
\centering
\vspace{-10em}
\includegraphics[width=0.9\textwidth]{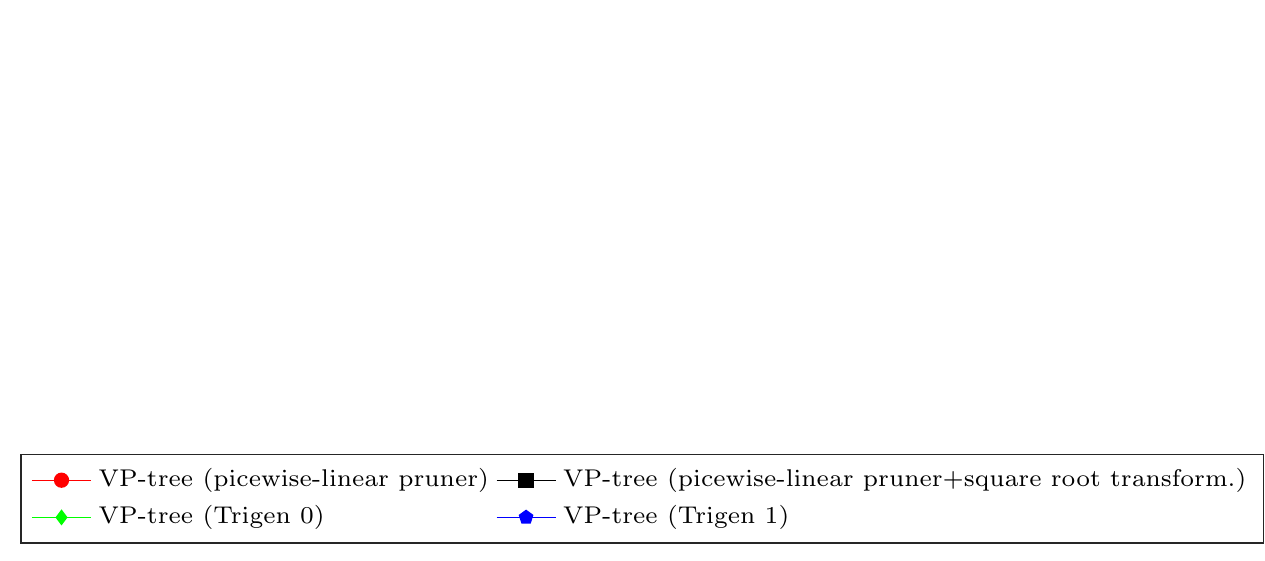}
\\
\subfloat[\scriptsize\label{PanelTrigen_eff_RCV8_KLdiv} RCV-8 (KL-div.)]{\includegraphics[width=0.320000\textwidth]{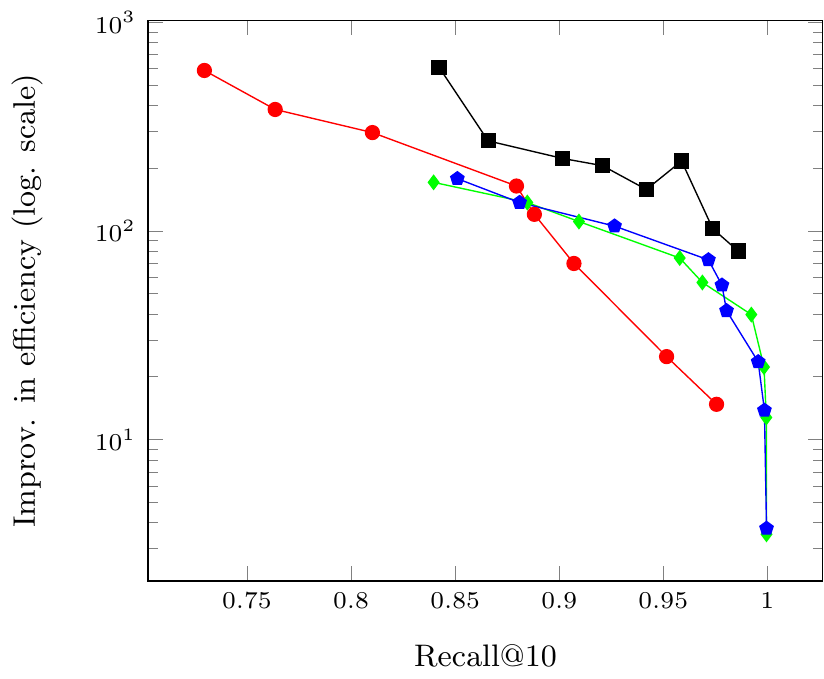}}
\subfloat[\scriptsize\label{PanelTrigen_eff_RCV8_RenyiDiv025} RCV-8 (R\'{e}nyi div. $\alpha=0.25$)]{\includegraphics[width=0.320000\textwidth]{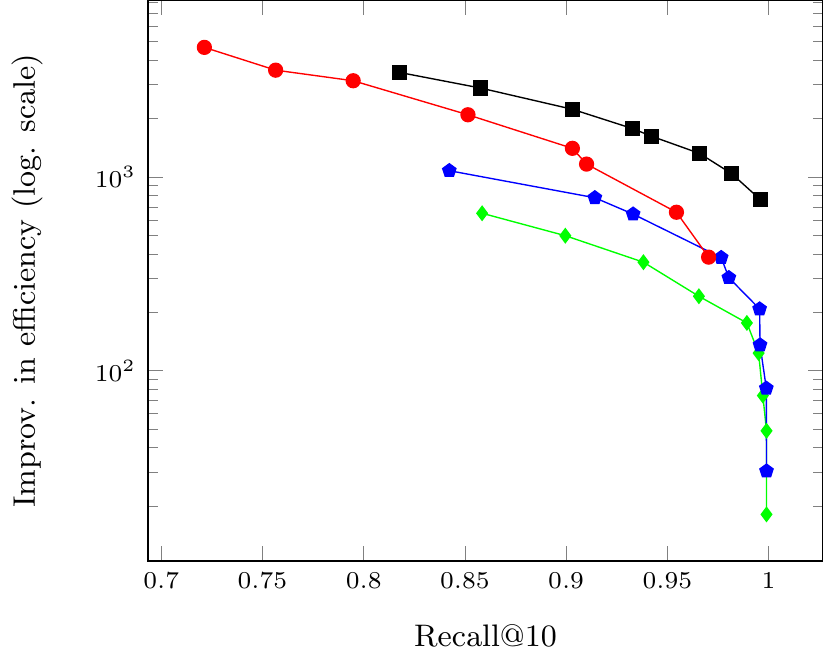}}
\subfloat[\scriptsize\label{PanelTrigen_eff_RCV8_RenyiDiv075} RCV-8 (R\'{e}nyi div. $\alpha=0.75$)]{\includegraphics[width=0.320000\textwidth]{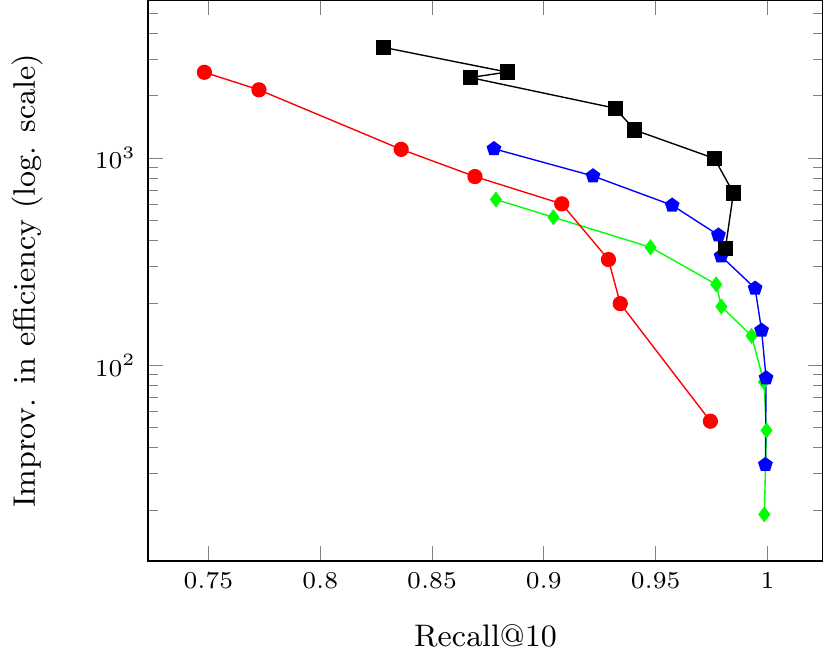}}
\\
\subfloat[\scriptsize\label{PanelTrigen_eff_Wiki8_Cosine} Wiki-8 (Cosine dist.)]{\includegraphics[width=0.320000\textwidth]{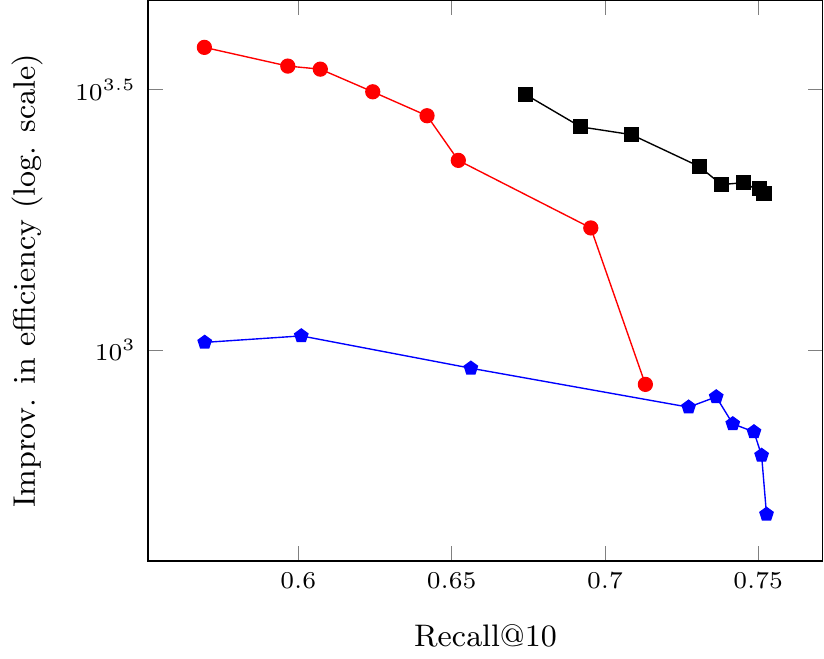}}
\subfloat[\scriptsize\label{PanelTrigen_eff_Wiki8_L2SQR} Wiki-8 ($L^2_2$)]{\includegraphics[width=0.320000\textwidth]{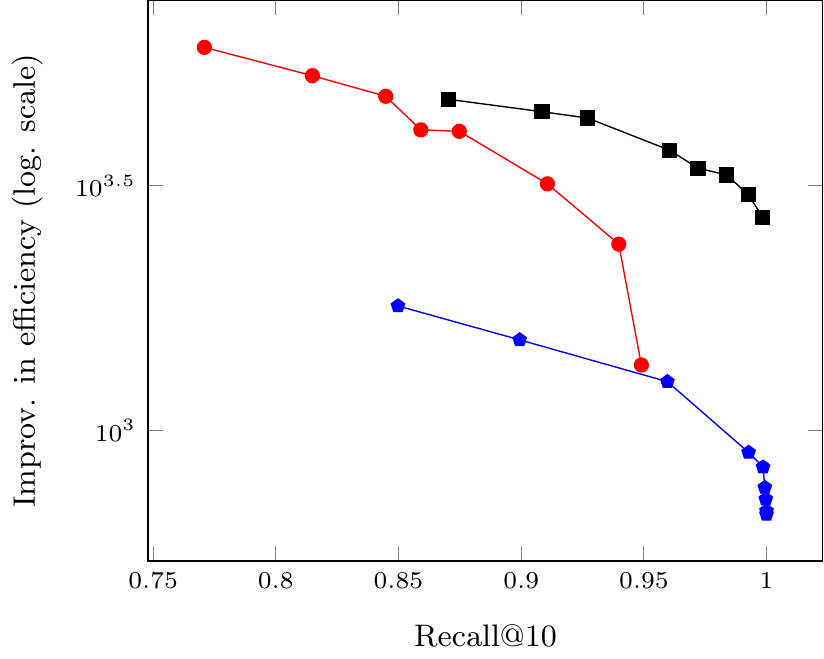}}
\subfloat[\scriptsize\label{PanelTrigen_eff_Wiki8_ItakuraSaito} Wiki-8 (Itakura-Saito)]{\includegraphics[width=0.320000\textwidth]{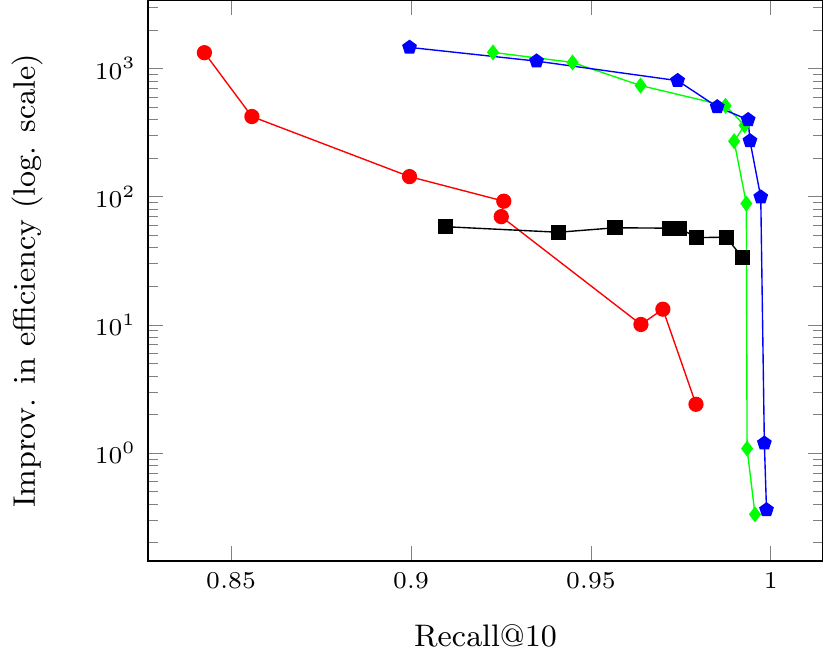}}
\\
\subfloat[\scriptsize\label{PanelTrigen_eff_Wiki8_KLdiv} Wiki-8 (KL-div.)]{\includegraphics[width=0.320000\textwidth]{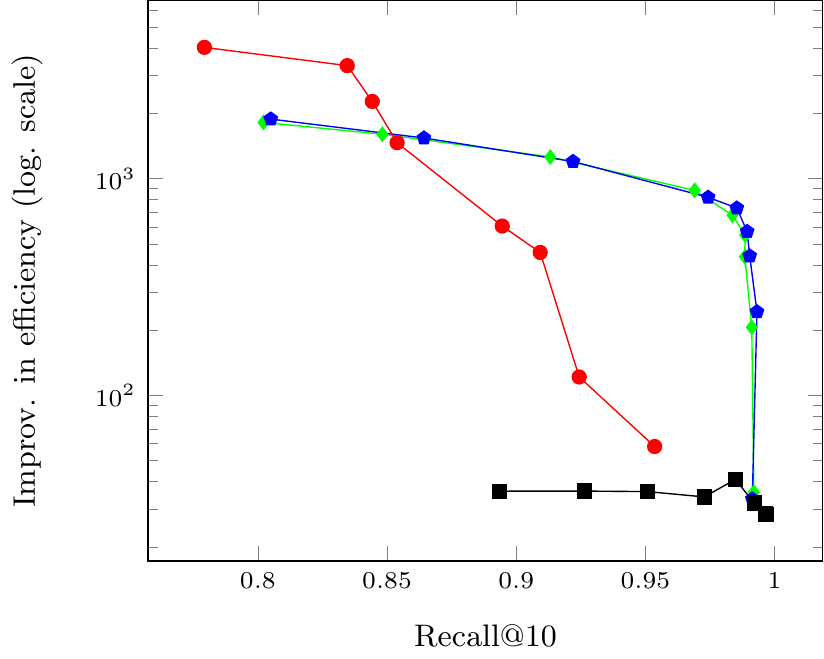}} 
\subfloat[\scriptsize\label{PanelTrigen_eff_Wiki8_RenyiDiv025} Wiki-8 (R\'{e}nyi div. $\alpha=0.25$)]{\includegraphics[width=0.320000\textwidth]{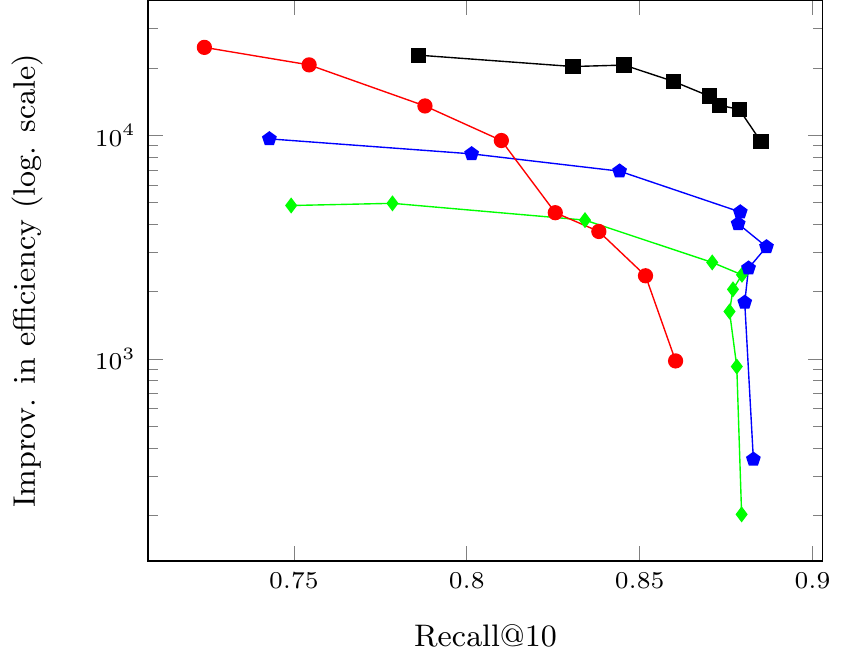}}
\subfloat[\scriptsize\label{PanelTrigen_eff_Wiki8_RenyiDiv075} Wiki-8 (R\'{e}nyi div. $\alpha=0.75$)]{\includegraphics[width=0.320000\textwidth]{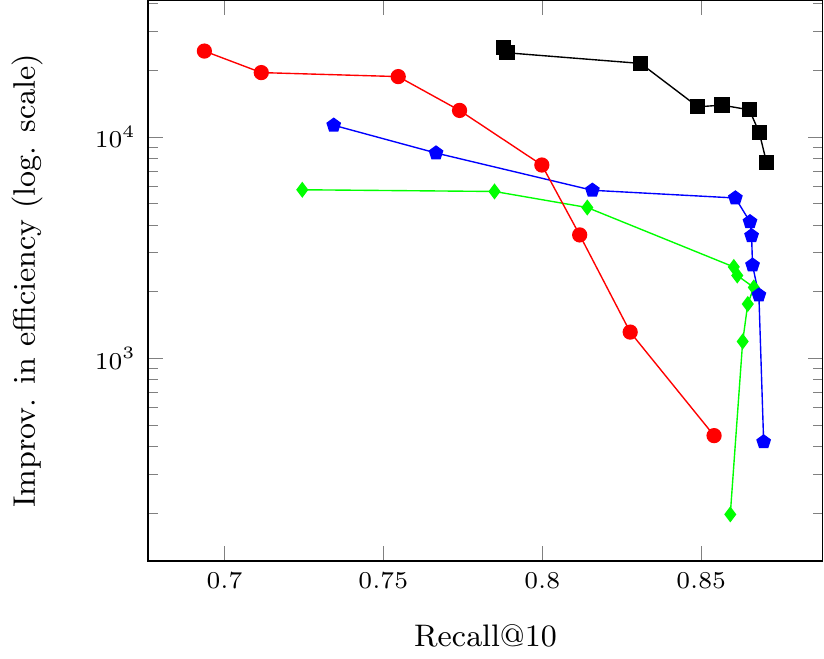}}
\\
\subfloat[\scriptsize\label{PanelTrigen_eff_RandHist8_KLdiv} RandHist-8 (KL-div.)]{\includegraphics[width=0.320000\textwidth]{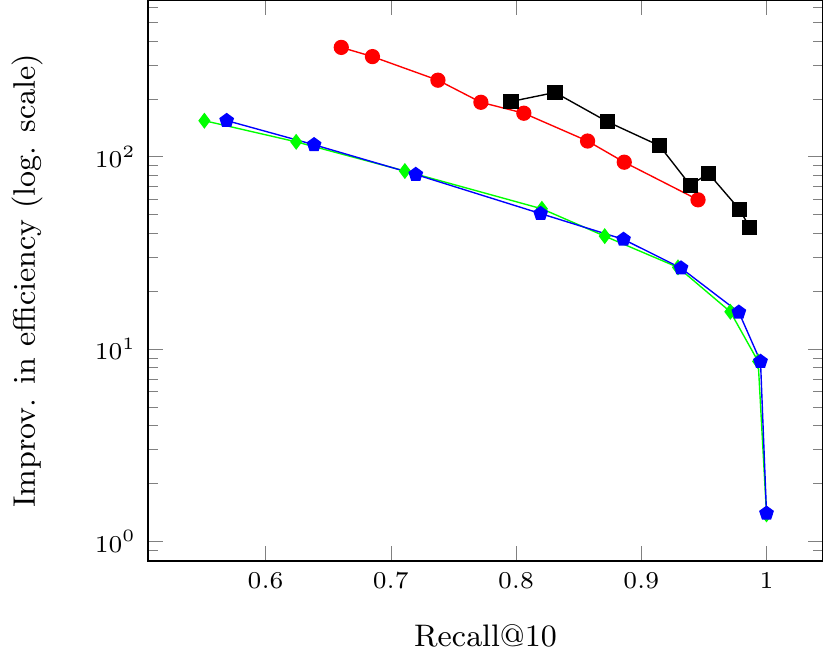}}
\subfloat[\scriptsize\label{PanelTrigen_eff_Wiki128_RenyiDiv025} Wiki-128 (R\'{e}nyi div. $\alpha=0.25$)]{\includegraphics[width=0.320000\textwidth]{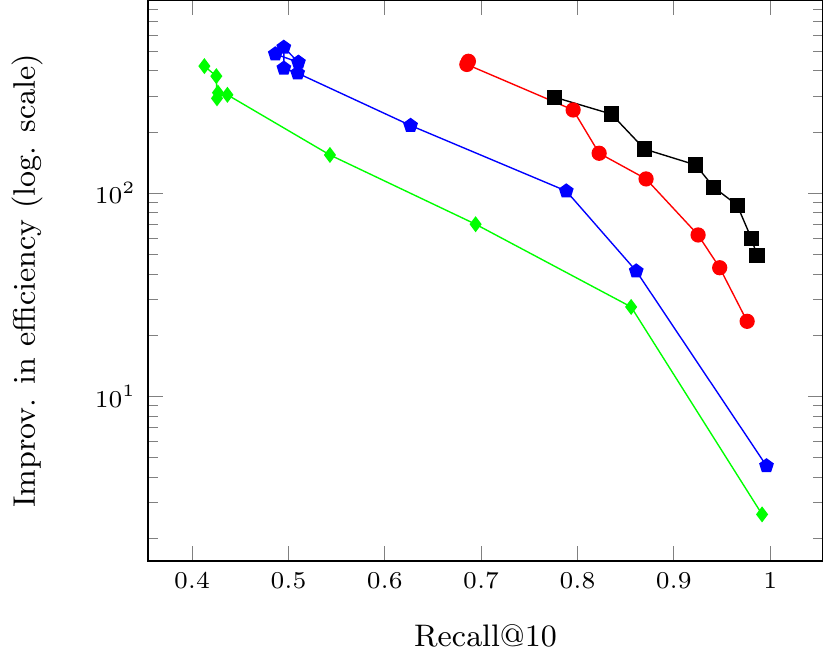}}
\subfloat[\scriptsize\label{PanelTrigen_eff_Wiki128_RenyiDiv2} Wiki-128 (R\'{e}nyi div. $\alpha=2$)]{\includegraphics[width=0.320000\textwidth]{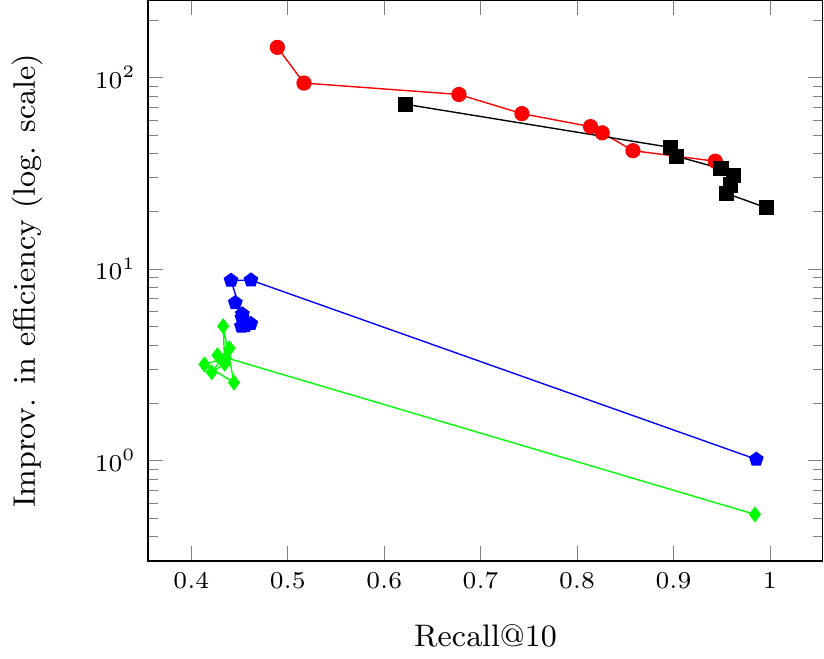}}
\\
\caption{\label{FigTrigen_eff} Improvement in efficiency vs recall for VP-tree based methods in 10-NN search. Best viewed in color.}
\end{figure}

\begin{figure}[t]
\centering
\vspace{-10em}
\includegraphics[width=0.9\textwidth]{images/trigen/legend_only_trigen.pdf}
\\
\subfloat[\scriptsize\label{PanelTrigen_dist_RCV8_KLdiv} RCV-8 (KL-div.)]{\includegraphics[width=0.320000\textwidth]{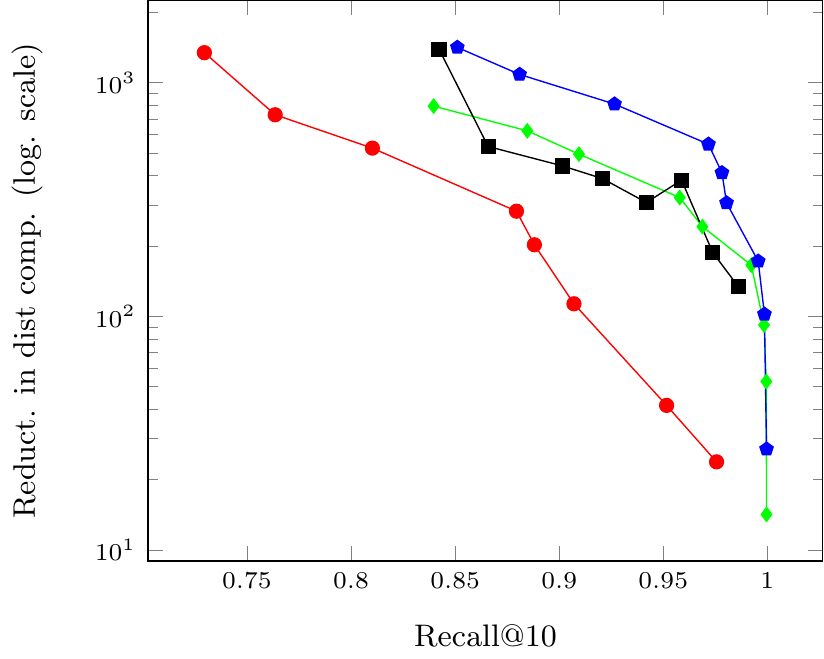}}
\subfloat[\scriptsize\label{PanelTrigen_dist_RCV8_RenyiDiv025} RCV-8 (R\'{e}nyi div. $\alpha=0.25$)]{\includegraphics[width=0.320000\textwidth]{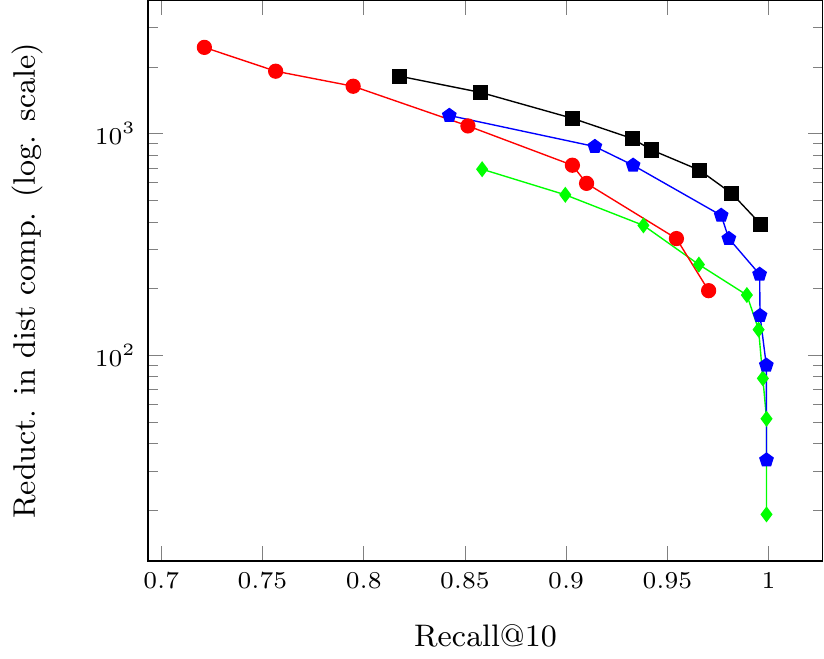}}
\subfloat[\scriptsize\label{PanelTrigen_dist_RCV8_RenyiDiv075} RCV-8 (R\'{e}nyi div. $\alpha=0.75$)]{\includegraphics[width=0.320000\textwidth]{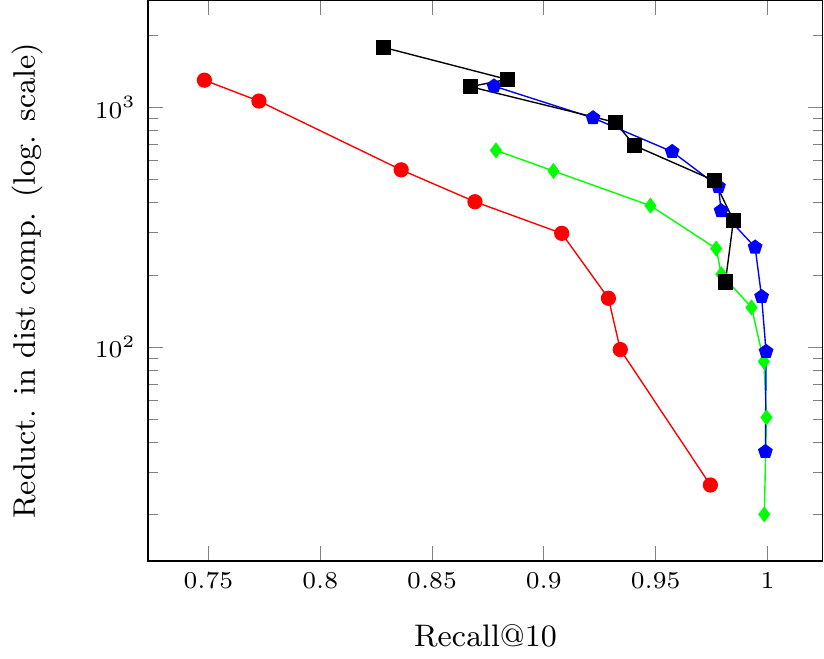}}
\\

\subfloat[\scriptsize\label{PanelTrigen_dist_Wiki8_Cosine} Wiki-8 (Cosine dist.)]{\includegraphics[width=0.320000\textwidth]{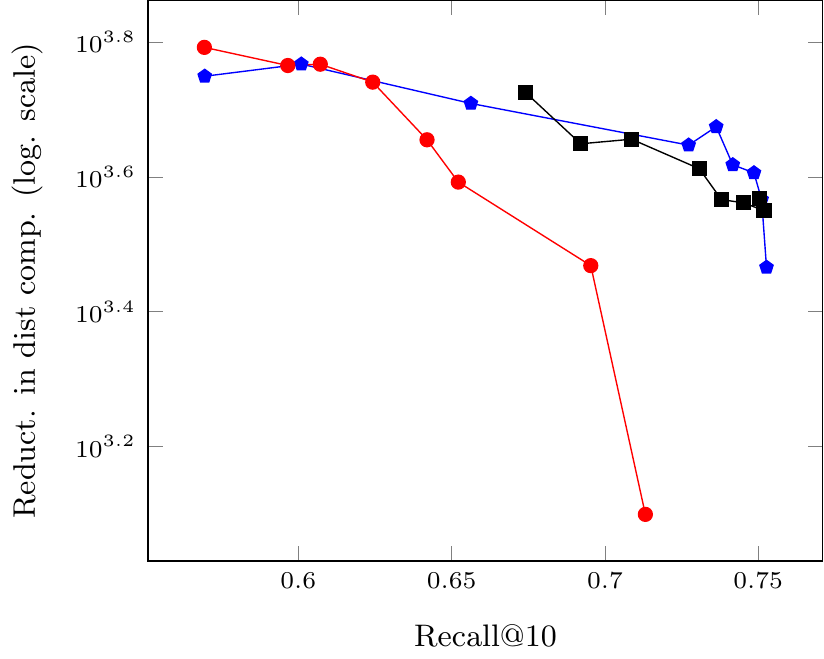}}
\subfloat[\scriptsize\label{PanelTrigen_dist_Wiki8_L2SQR} Wiki-8 ($L^2_2$)]{\includegraphics[width=0.320000\textwidth]{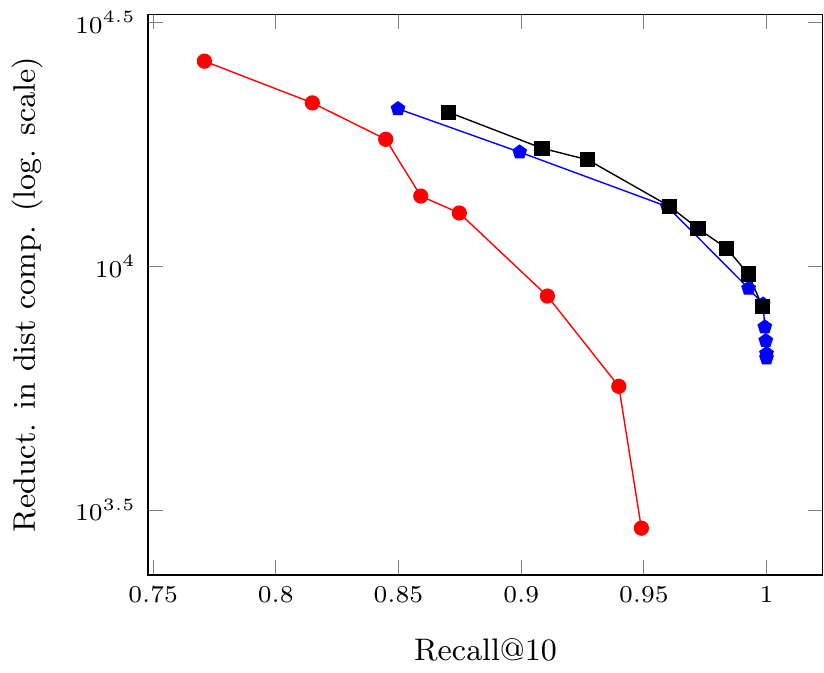}}
\subfloat[\scriptsize\label{PanelTrigen_dist_Wiki8_ItakuraSaito} Wiki-8 (Itakura-Saito)]{\includegraphics[width=0.320000\textwidth]{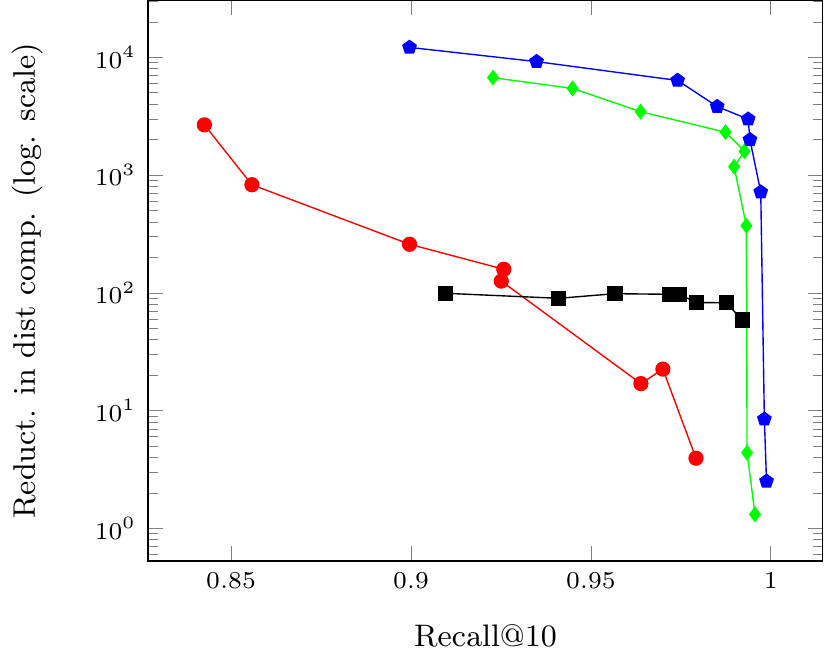}}
\\
\subfloat[\scriptsize\label{PanelTrigen_dist_Wiki8_KLdiv} Wiki-8 (KL-div.)]{\includegraphics[width=0.320000\textwidth]{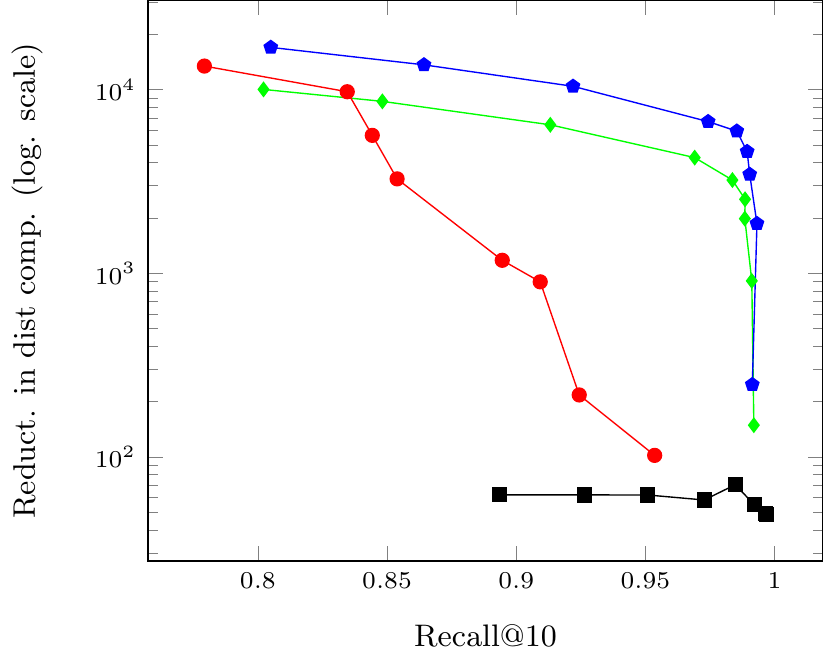}} 
\subfloat[\scriptsize\label{PanelTrigen_dist_Wiki8_RenyiDiv025} Wiki-8 (R\'{e}nyi div. $\alpha=0.25$)]{\includegraphics[width=0.320000\textwidth]{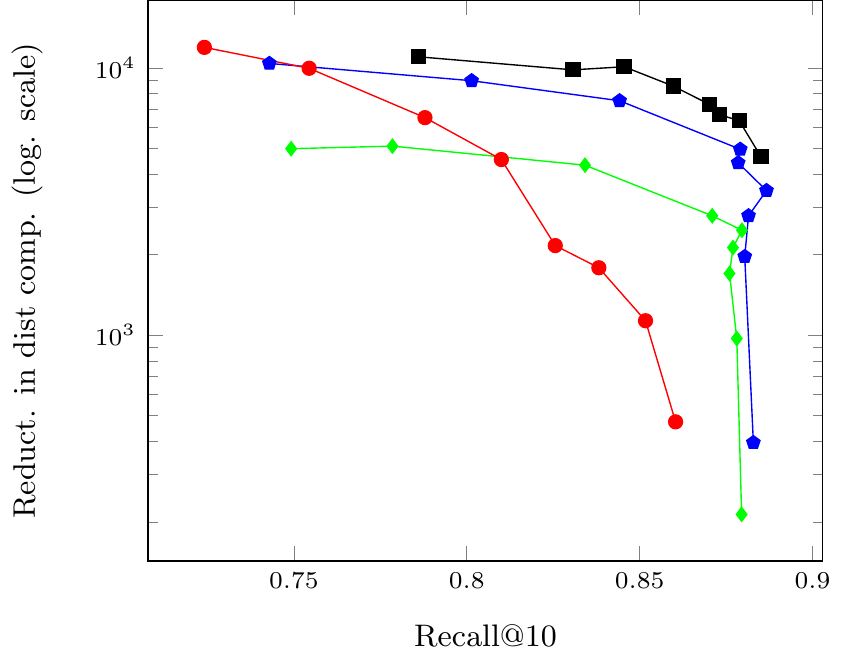}}
\subfloat[\scriptsize\label{PanelTrigen_dist_Wiki8_RenyiDiv075} Wiki-8 (R\'{e}nyi div. $\alpha=0.75$)]{\includegraphics[width=0.320000\textwidth]{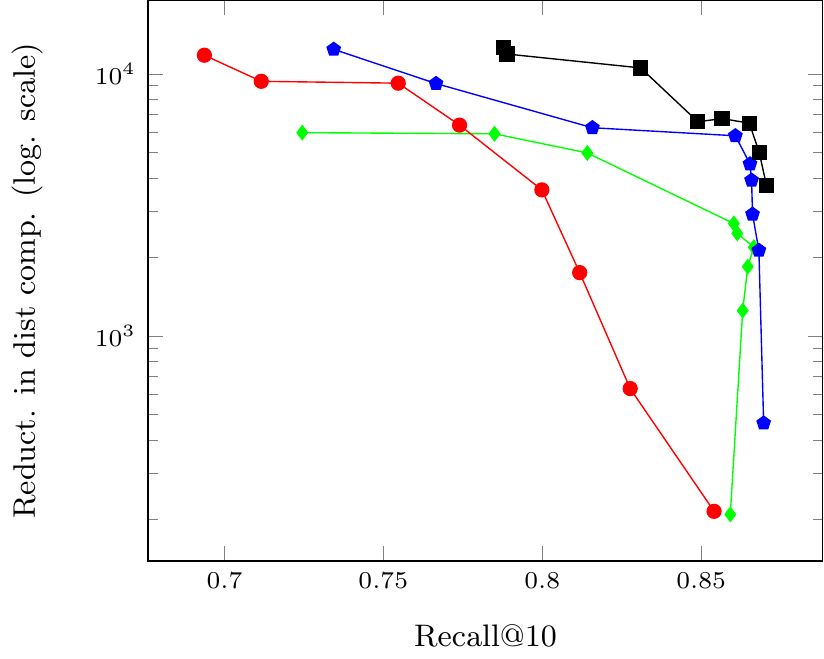}}
\\
\subfloat[\scriptsize\label{PanelTrigen_dist_RandHist8_KLdiv} RandHist-8 (KL-div.)]{\includegraphics[width=0.320000\textwidth]{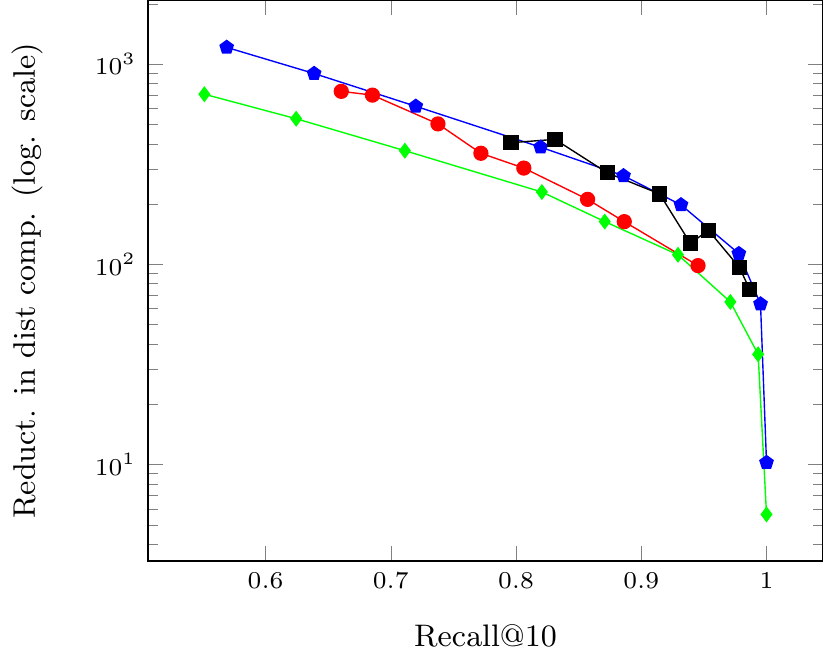}}
\subfloat[\scriptsize\label{PanelTrigen_dist_Wiki128_RenyiDiv025} Wiki-128 (R\'{e}nyi div. $\alpha=0.25$)]{\includegraphics[width=0.320000\textwidth]{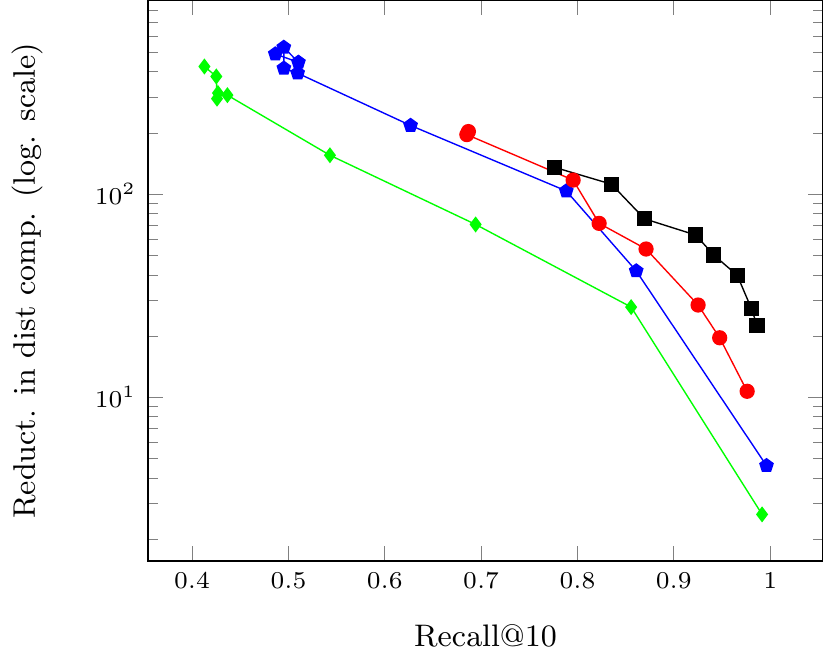}}
\subfloat[\scriptsize\label{PanelTrigen_dist_Wiki128_RenyiDiv2} Wiki-128 (R\'{e}nyi div. $\alpha=2$)]{\includegraphics[width=0.320000\textwidth]{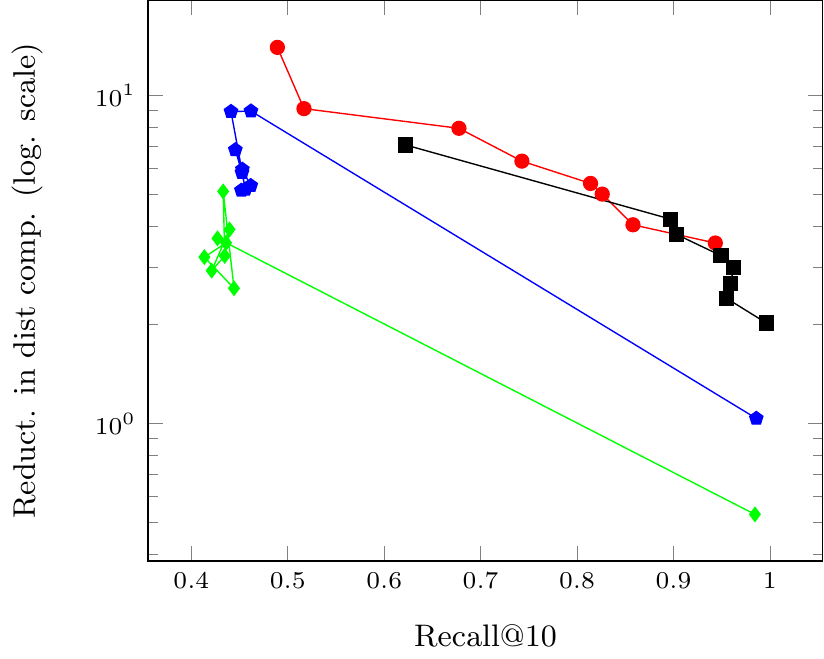}}
\\
\caption{\label{FigTrigen_dist} Reduction in the number of distance computations vs recall for VP-tree based methods in 10-NN search. Best viewed in color.}
\end{figure}

%% file: conclusion.tex
\section{Conclusion}

We carry out the first comparison of two generic pruning approaches
for non-metric data. 
Our approach is comprehensive and involves 40 combinations of 
data sets and distances, which 
cannot be handled by a classic metric-space access method.
We extend TriGen to the case of non-symmetric distances
and  demonstrate that VP-tree with a data-adapted
pruning rule can enable accurate non-metric \knn search for data of moderate dimensionality
by using the modified TriGen, the piecewise linear approximation
of the metric pruning rule, or by the hybrid approach.
In that, we find that this hybrid approach is often more effective
than either of the pruning rules.
Our software is publicly available:
NMSLIB branch \ttt{nmslib4a\_bigger\_reruns},
search method \ttt{vptree\_trigen}.\footnote{ \url{https://github.com/nmslib/nmslib/tree/nmslib4a_bigger_reruns}}

\paragraph{Acknowledgments}
This work was done while Leonid Boytsov was a PhD student at CMU.
Authors gratefully acknowledge the support by the NSF grant \#1618159